\documentstyle[12pt]{article}	
\catcode`\@=11 \@ifundefined{tencircw}{}{}
\catcode`\@=12

\textheight 9in  \topmargin -.4in   \def\baselinestretch{1.2}
\def\Eq{\begin{equation}}	\def\End{\end{equation}}
\def\Eqa{\begin{eqnarray}}	\def\Enda{\end{eqnarray}}
\def\Endl#1{\label{#1} \End}	\def\Endla#1{\label{#1} \Enda}
\def\puteq#1{eq.~(\ref{#1})}	
\def\d{{\rm d}}		\def\to{\!\rightarrow\!}\def\ord#1{{\cal O}(#1)}
\def\etal{{\it et.al.}}	\def\ie{{\it i.e.}}	\def\eg{{\it e.g.}}
\def\a{\alpha} \def\b{\beta} \def\g{\gamma} \def\p{\phi} \def\l{\lambda}
\def\slp{(6\l\p)}	\def\mub{\bar\mu}
\def\bM{\bar M}		\def\bm{\bar m}
\def\abg#1#2#3{\quad\ord{\a^{#1}\b^{#2}\g^{#3}}}
\def\ab#1#2{\abg{#1}{#2}{0}}
\def\frac#1#2{{\textstyle {#1 \over #2}}}
\def\mfrac#1#2{{\textstyle {- \atop }\!{#1 \over #2}}}
\def\sip{T \sum_n \int {\d^3 {\bf p} \over (2\pi)^3}}

\newcounter{xpos} \newcounter{ypos}
\def\Bp{\begin{picture}}	\def\Ep{\end{picture}}
\def\core#1#2#3#4#5#6{\raisebox{#1pt}{\Bp(#2,#3) \setcounter{xpos}{#4}
  \setcounter{ypos}{#5} #6 \Ep}}
\def\gtad#1#2#3#4{\hspace{2pt} \core{-17}{#1}{#2}{#3}{20}{
  \put(#3,0){\line(0,1){10}} #4} \hspace{2pt}}
\def\vac#1#2#3#4#5{\mfrac1{#1} \hspace{2pt} \core{ -7}{#2}{#3}{#4}{10}{#5}
  \hspace{2pt}}
\def\tad#1#2#3#4#5{\mfrac1{#1} \gtad{#2}{#3}{#4}{#5}}
\def\Mr{\addtocounter{xpos}{ 20}}  \def\Ml{\addtocounter{xpos}{-20}}
\def\Mu{\addtocounter{ypos}{ 20}}  \def\Md{\addtocounter{ypos}{-20}}
\def\MR{\addtocounter{xpos}{ 15}}  \def\ML{\addtocounter{xpos}{-15}}
\def\MU{\addtocounter{ypos}{ 15}}  \def\MD{\addtocounter{ypos}{-15}}
\def\Cf{\put(\value{xpos},\value{ypos}){\circle*{3}}}		
\def\Ca{\put(\value{xpos},\value{ypos}){\circle{20}}}		
\def\Cai{\Ca \put(\value{xpos},\value{ypos}){\circle{16}}}	
\def\Cz{\Ca\Cf}		\def\Czi{\Cai\Cf}			
\def\Cn{\put(\value{xpos},\value{ypos}){\circle{10}}}		
\def\Cni{\Cn \put(\value{xpos},\value{ypos}){\circle{ 7}}}
\def\thk#1{\thicklines #1 \thinlines}
\def\blot{\rule{4pt}{4pt}}
\def\tadlead{\mfrac12 \hspace{2pt} \core{-10}{10}{20}{5}{15}{
    \put(5,0){\line(0,1){10}}\Cn} \hspace{2pt}}

\begin{document}
\begin{titlepage}
\begin{center}
April, 1993	\hfill CALT-68-1858\\
		\hfill HUTP-93-A011\\
		\hfill EFI-93-22\\
\vskip.5in {\large\bf
Resummation Methods at Finite Temperature:\\The Tadpole Way%
}\footnote{Work supported in part by the U.S. Dept. of Energy under
Contracts DEAC-03-81ER40050 and DEFG-02-90ER40560.}\\
\vskip.3in
{
  {\bf C. Glenn Boyd}\footnote{Email: \tt boyd@rabi.uchicago.edu}\\
  \vskip.2cm {\it Enrico Fermi Institute, 5640 Ellis Ave., Chicago IL 60637}
\vskip.3cm
  {\bf David E. Brahm}\footnote{Email: \tt brahm@theory3.caltech.edu}\\
  \vskip.2cm {\it Caltech 452-48, Pasadena CA 91125}
\vskip.3cm  and \vskip.3cm
  {\bf Stephen D.H. Hsu}\footnote{Junior Fellow, Harvard Society of
     Fellows.  Email: \tt hsu@hsunext.harvard.edu }
  \vskip.2cm {\it Lyman Laboratory of Physics, Harvard University,
     Cambridge MA 02138}}
\end{center}

\vskip.5in
\begin{abstract}
We examine several resummation methods for computing higher order
corrections to the finite temperature effective potential, in the context
of a scalar $\p^4$ theory.  We show by explicit calculation to four loops
that dressing the propagator, not the vertex, of the one-loop tadpole
correctly counts ``daisy'' and ``super-daisy'' diagrams.
\end{abstract}
\end{titlepage}
\setcounter{footnote}{0}


\section{Introduction: Resummation of Daisies}

Recent interest in the electroweak phase transition (EWPT) has led to
attempts to improve the finite temperature effective potential by
resummation of leading infrared divergent graphs [1-7].
There is some controversy over the correct resummation procedure,
particularly how ``daisy'' and ``superdaisy'' \cite{dj} graphs are
accounted for by dressing the propagator and/or vertex in 1-loop tadpole
graphs.  Espinosa, Quiros and Zwirner \cite{eqz1,eqz2} advocate dressing
both propagator and vertex (``full dressing''), while we claim \cite{us2}
that dressing the propagator alone (``partial dressing'') accurately counts
higher-loop graphs.  In this paper we show by explicit calculation to four
loops in a scalar theory that partial dressing reproduces the correct
combinatorics, while full dressing overcounts an infinite set of diagrams.

We also discuss other methods in the literature.  Arnold and Espinosa
\cite{deriv} have reported that resummation corrections make the EWPT more
strongly first order.  We verify that their counting scheme, applied to the
scalar theory, is equivalent to partial dressing; and unlike either
superdaisy resummation scheme, it handles overlapping momenta correctly.

In Sec.~2, after introducing our notation, we compare full to partial
dressing (graphically and algebraically) in a region of parameter space
where trilinear couplings are small, and calculate the higher-loop diagrams
explicitly.  Trilinear couplings are considered in more detail in Sec.~3,
some earlier approximations are eliminated, and questions of overlapping
momenta are addressed.  Our result for the effective potential is then
presented.  In Sec.~4 we examine the Cornwall, Jackiw and Tomboulis
procedure \cite{acp,cjt}, Arnold
and Espinosa's loop expansion \cite{deriv}, and the 2-point method of
Buchmuller \etal\ \cite{buch}, and suggest a hybrid method with the best
features of the others.  We summarize our findings in Sec.~5.

\section{Vacuum or Tadpole?}

Consider a real scalar field theory with tree-level potential $V_0 =
\frac\l4 \p^4 - \frac{\mu^2}2 \p^2$.  The effective potential is given by
minus the sum of all vacuum-to-vacuum graphs; the 1-loop contribution is
  \Eq \vac{2}{20}{20}{10}{\Ca} = \vac{2}{10}{15}{5}{\Cn} +
    \vac{2}{20}{20}{10}{\Cz} = \left({-\pi^2 T^4\over90} + {T^2 m^2\over24}
    + \cdots \right) - {T m^3\over12\pi} \Endl{vac}
We have separated the $n\!\ne\!0$ modes (small loop) and the $n\!=\!0$ mode
(dotted loop), displayed symmetry factors explicitly, and kept only terms
$\ord{T}$ or higher.  Here $m^2 = V_0'' = 3\l\p^2-\mu^2$, and $n$ is the
Matsubara frequency index.  Over regions of $\p$ where $m\to0$, infrared
divergences appearing in the zero-mode contribution must be compensated by
including higher-loop ``daisy'' and ``superdaisy'' diagrams \cite{dj},
which give the scalar an effective ``plasma mass.''

An alternative approach \cite{lee} is to calculate the derivative $V'(\p)$,
given by the sum of all tadpole graphs, and then integrate with respect to
$\p$.  The tadpoles are given correctly, including symmetry factors, by
attaching a ($p\!=\!0$ truncated) external line to each part of each vacuum
graph, \eg
  \Eq {\d\over\d\p} \left[ \mfrac18 \hspace{2pt}
    \core{-17}{20}{40}{10}{10}{\Ca\Mu\Ca} \hspace{3pt} \right] = \left[
    \mfrac14 \hspace{2pt} \core{-22}{20}{50}{10}{20}{
      \put(10,0){\line(0,1){10}}\Ca\Mu\Ca} \hspace{3pt} \right] \Endl{tad}
When we discuss a method from the literature which uses vacuum graphs (\eg\
\cite{eqz1,eqz2,deriv,acp,cjt}), we will usually convert it to tadpoles
(take $\d/\d\p$) to facilitate a comparison.

While hard thermal loops (daisies) can be included by shifting the mass
with a temperature dependent term in either vacuum or tadpole graphs,
higher order corrections (superdaisies) require a field dependent mass
shift which gives different results if inserted into vacuum rather than
tadpole diagrams.  There has been confusion in the literature over which
method properly incorporates the important higher loop superdaisy diagrams.
It is known, for example, that a simple shift $m^2 \to m^2 + \Pi(\p,T)$ in
the one loop contribution to the effective potential \puteq{vac}, where
$\Pi$ is the scalar self-energy, results in an overcounting of the 2-loop
figure-eight vacuum diagram on the left side of \puteq{tad}\footnote{In
\cite{eqz1}, when their equation (33) is substituted in (21) and expanded,
the result for $V$ includes a term $3\l T^2 m^2/32\pi^2$, which is twice
the correct result.}.  Shifting $m^2$ in the vacuum diagram is equivalent
to dressing the propagator and 3-point vertex of the 1-loop tadpole, so
this ``full dressing'' overcounts the 2-loop figure-eight tadpole on the
right side of \puteq{tad} \cite{us2,linde}.  We will show, by explicit
calculation to 4 loops, that dressing only the propagator in a 1-loop
tadpole (``partial dressing'') correctly counts the relevant graphs.

\subsection{Dressing Up}

In both full and partial dressing procedures, the propagator is first
improved by solving a gap equation:\footnote{In all diagrammatic mass gap
equations, we display only the 1PI diagrams from the usual infinite series.}
\Eq \Bp(32,3) \put(0,0){\line(1,0){32}} \put(0,3){\line(1,0){32}} \Ep \;=\;
    \Bp(32,1) \put(0,1){\line(1,0){32}} \Ep \;+\;
    \core{0}{32}{20}{16}{10}{\Cai \put(0,0){\line(1,0){32}}} \;+\;
    \core{-10}{32}{20}{16}{10}{\Cai \put(0,10){\line(1,0){6}}
      \put(26,10){\line(1,0){6}}}  \qquad
  M^2 = m^2 - \frac12 c_4 I_1(M^2) - \frac12 c_3^2 I_2(M^2) \Endl{mgap}
$c_3$ and $c_4$ are the 3- and 4-point vertices, respectively, with $\d
c_3/\d\p = c_4$.  We have defined
  \Eqa I_0(m^2) &\equiv& -\sip \ln\left[(2\pi n T)^2 + p^2 + m^2 \right],
      \nonumber \\
    I_j(m^2) &\equiv& \phantom{-}\sip \left[(2\pi n T)^2 + p^2 + m^2
      \right]^{-j} \quad (j>0) \Enda
The dressed vertex is found by differentiating the mass gap equation and
solving for the improved three point function:
\Eq \core{-5}{32}{18}{0}{0}{\put(16,0){\line(0,1){16}} \put(14,14){\blot}
      \put(0,15){\line(1,0){32}} \put(0,17){\line(1,0){32}}} \;=\;
    \core{-5}{32}{18}{0}{0}{\put(16,0){\line(0,1){16}}
      \put(0,15){\line(1,0){32}} \put(0,17){\line(1,0){32}}} \;+\;
    \core{-10}{32}{29}{16}{17}{\Cai \put(16,0){\line(0,1){7}}
      \put(14,6){\blot} \put(0,27){\line(1,0){32}}
      \put(0,29){\line(1,0){32}}} \;+\;
    2 \; \core{-10}{32}{26}{16}{16}{\Cai \put(0,15){\line(1,0){6}}
      \put(0,17){\line(1,0){6}} \put(26,15){\line(1,0){6}}
      \put(26,17){\line(1,0){6}} \put(6,0){\line(0,1){16}}} \;+\;
    \core{-10}{32}{27}{16}{17}{\Cai \put(0,16){\line(1,0){6}}
      \put(0,18){\line(1,0){6}} \put(26,16){\line(1,0){6}}
      \put(26,18){\line(1,0){6}} \put(16,0){\line(0,1){7}}
      \put(14,6){\blot}} \Endl{vga}
  \Eq C_3 \equiv  - {\d M^2 \over \d\p} = c_3 + \frac12 C_3 c_4 I_2(M^2) +
    c_3 c_4 I_2(M^2) +  C_3 c_3^2 I_3(M^2) \Endl{vgap}
Note for later reference that the sub-diagram
  $\core{-11}{30}{31}{15}{16}{\Ca \put(15,0){\line(0,1){6}}
  \put(0,26){\line(1,0){30}}}$
appears three times in the improved vertex, once from the penultimate
term in \puteq{mgap} and twice from the last term.

The prescription in \cite{us2} is to equate $V'$ with a one loop
improved propagator tadpole (partial dressing)
\Eq V'_{pd} = \gtad{20}{30}{10}{\Cai} =
    \gtad{10}{20}{5}{\Md\MU\Cni} + \gtad{20}{30}{10}{\Czi} =
    \gtad{10}{20}{5}{\Md\MU\Cn} + \gtad{10}{30}{5}{\Md\MU\Cn\MU\MU\Md\Cn} +
    \gtad{20}{40}{10}{\Md\MU\Cn\MU\Cz} + \gtad{20}{30}{10}{\Cz} +
    \gtad{20}{40}{10}{\Cz\MU\Cn} + \gtad{20}{50}{10}{\Cz\Mu\Cz} +
    \gtad{20}{70}{10}{\Cz\Mu\Cz\Mu\Cz} + \cdots \End
This contrasts with the procedure in \cite{eqz1}, which consists of
substituting $M^2$ into the one loop potential $V = \mfrac12 I_0(M^2)$,
expanding and replacing $M^2$ by $m^2$ in all terms except the cubic $M^3$.
Since the $M^3$ term arises precisely from the zero mode in the
the one loop tadpole, differentiating this expression by $\p$ and
expanding the vertex gives
\Eqa V'_{fd} = \gtad{10}{20}{5}{\Md\MU\Cn} + \gtad{20}{30}{10}{\Czi
    \put(8,9){\blot}} &=&
  \gtad{10}{20}{5}{\Md\MU\Cn} + \gtad{20}{30}{10}{\Czi} +
  \gtad{20}{50}{10}{\Cai\Mu\Czi \put(8,9){\blot}} +
  3\; \gtad{20}{30}{10}{\Cai \put(9,10){\line(0,1){20}}
    \put(11,10){\line(0,1){20}} \put(14,20){\circle*{3}}} +
  2\; \gtad{20}{30}{10}{\Cai \put(8,9){\blot} \put(0,19){\line(1,0){20}}
    \put(0,21){\line(1,0){20}} \put(10,24){\circle*{3}}} \nonumber \\
  &=& \gtad{10}{20}{5}{\Md\MU\Cn} + \gtad{20}{40}{10}{\Md\MU\Cn\MU\Cz}
       + \gtad{20}{30}{10}{\Cz} + \gtad{20}{40}{10}{\Cz\MU\Cn} +
    2\; \gtad{20}{50}{10}{\Cz\Mu\Cz} +
    3\; \gtad{20}{70}{10}{\Cz\Mu\Cz\Mu\Cz} + \cdots \Endla{fdress}
A dot inside a loop means only zero modes are contained in that loop
variable, although non-zero modes can run through shared propagators.

Although we have expanded the improved vertex and propagators graphically,
the algebraic expansion is equally simple by repeated use of the
recursion relation \puteq{vgap} and
  \Eq \frac{c_3}2 [I_1(M^2)] = \frac{c_3}2 [I_1(m^2) + \frac12 c_4 I_2(m^2)
    I_1(M^2) + \frac14 c_4^2 I_1^2(M^2) I_3(m^2) + \frac18 c_4^3 I_1^3(M^2)
    I_4(m^2) + \cdots]  \Endl{propr}
  \Eq \gtad{20}{30}{10}{\Cai} = \gtad{20}{30}{10}{\Ca} +
    \gtad{20}{50}{10}{\Ca\Mu\Cai} + \gtad{60}{30}{30}{\Ca\Mr\Cai\Ml\Ml\Cai}
    + \gtad{60}{50}{30}{\Ca\Mr\Cai\Ml\Ml\Cai\Mr\Mu\Cai} + \cdots \End
where the ellipses refers to terms with powers of $c_3$ and higher powers
of $c_4$.  The result can then be compared to a Feynman graph computation
of the one point function to examine the validity of the two methods. In
practice, gap equation solutions and loop integrals can only be
approximated.  However, as long as the same approximations are made in the
Feynman diagram expansion, the results can still be consistently and
explicitly compared.

\subsection{Rules of the Game}

We now calculate the algebraic expressions and the Feynman diagrams for a
real scalar theory, under some simplifying rules.  All diagrams are
preceded by an overall minus sign, to give $V$ or its derivatives, and a
symmetry factor.  The vertices are $c_3=-V_0''' = -6\l\p$ and $c_4=-V_0''''
= -6\l$.  We ignore for now the ellipsis in \puteq{vac}; we will discuss
the missing terms in Sec.~3.  Temperature dependent parts of loop
integrals are then given by Table~\ref{tab1} (our combinatoric analysis
will not hinge on zero temperature results, or on renormalization
prescriptions).  The leading-order result
comes from a single hard (non-zero mode) thermal loop
  \Eq \tadlead = \slp {T^2\over24} \Endl{lead}
Note that in this approximation (which we re-examine in Sec.~3),
higher-loop diagrams made by attaching bubbles to this one vanish, due to
the zeros in the first line of Table~\ref{tab1}.

  \begin{table} \centering \begin{tabular}{|r|ccccc|} \hline
  integral: & $I_0$ & $I_1$ & $I_2$ & $I_3$ & $I_4$ \\ \hline
  non-zero modes: & $\left(\frac{\pi^2 T^4}{45} \!-\! \frac{T^2 m^2}{12}
    \right)$ & $\frac{T^2}{12}$ & 0 & 0 & 0 \rule[-8pt]{0pt}{23pt}\\
  zero modes: & $\frac{T m^3}{6\pi}$ & $\frac{-Tm}{4\pi}$ & $\frac{T}{8\pi
    m}$ & $\frac{T}{32\pi m^3}$ & $\frac{T}{64\pi m^5}$
    \rule[-8pt]{0pt}{23pt}\\ \hline
  \end{tabular} \caption{Loop integrals} \label{tab1} \end{table}

Diagrams are categorized as being $\ord{\a^{j_\a} \b^{j_\b} \g^{j_\g}}$
with respect to \puteq{lead}, where \cite{eqz1}
  \Eq \a \equiv \lambda T^2/m^2 \approx 1, \qquad
      \b \equiv \lambda T/m < 1, \qquad
      \g \equiv \p^2/T^2 \ll 1 \End
in the region of interest to us.  To keep things tractable we will first
look only at $\ord{\g^0}$, meaning only one 3-point vertex; higher orders
are discussed in Sec.~3.  A scheme is called ``accurate to $\ord{\b^{j}}$''
if it correctly reproduces all diagrams with $j_\b \le j$ and
$j_\g=0$.\footnote{In \cite{eqz1} diagrams are compared instead to the
leading zero-mode loop, so their ``$\ord{\bar\b}$'' corresponds to our
$\ord{\b^2}$.}  In \cite{us2} we call $\ord{\b}$ ``daisy order,'' and
$\ord{\b^2}$ ``super-daisy order.''  While the schemes described in this
paper are not accurate to $\ord{\b^{3}}$, we find it instructive here to
analyze ``daisy-type'' graphs to $\ord{\b^{4}}$.  For illustrative
purposes, we will display terms and diagrams only to $\ord{\l^4}$, which
for $j_\g=0$ means 4 loops.

\subsection{Partial- and Full-Dressing Results}

To the order we are working, the gap equation is
\Eq \Bp(32,3) \put(0,0){\line(1,0){32}} \put(0,3){\line(1,0){32}} \Ep \;=\;
    \Bp(32,1) \put(0,1){\line(1,0){32}} \Ep \;+\;
    \Bp(32,10) \put(0,0){\line(1,0){32}} \put(16,6){\circle{10}} \Ep \;+\;
    \core{0}{32}{20}{16}{10}{\Czi \put(0,0){\line(1,0){32}}} \;+\;
    \core{-10}{32}{20}{16}{10}{\Czi \put(0,10){\line(1,0){6}}
      \put(26,10){\line(1,0){6}}}  \qquad
  M^2 = m^2 + {\l T^2\over4} - {3\l TM\over4\pi} - {9\l^2 \p^2 T\over
    4\pi M} \Endl{gap}
Though the last term is $\ord{\a\g}$ compared to the previous one, we
retain it because their {\it derivatives\/} are the same order.  We need
not dress the non-zero-mode loop at this level of approximation, as
discussed below \puteq{lead}.

The solution of the gap equation, expanded to $\ord{\l^3 \g^0}$, is
  \Eq M = m + {\l T^2\over 8m} - {3\l T\over 8\pi} - {\l^2 T^4
    \over 128 m^3} + {9\l^2 T^2\over 128\pi^2 m} + {\l^3 T^6\over 1024 m^5}
    - {9\l^3 T^4 \over 1024\pi^2 m^3} + \ord{\l^4,\g^1} \Endl{Mimp}
and the improved 3-point coupling is
  \Eq -{\d M^2 \over \d\p} = -\slp \left[ 1 + \left\{ - {9\l T\over 8\pi m}
    + {9\l^2 T^3\over 64\pi m^3} - {27\l^3 T^5 \over 1024\pi m^5} +
    {81\l^3 T^3\over 1024\pi^3 m^3} \right\} + \ord{\l^4,\g^1} \right]
    \Endl{cimp}
As noted earlier, exactly $1/3$ of the expression in curly brackets arises
from the penultimate term in \puteq{gap}, and $2/3$ from the last term.

The partially-dressed 1-loop tadpole is then
  \Eqa \lefteqn{\tad{2}{20}{30}{10}{\Cai} = \tadlead + \slp {-TM\over8\pi}
    = \tadlead + \slp\times} \label{tm} \\
  && \left[ {-Tm\over8\pi} - {\l T^3\over64\pi m} + {3\l T^2
    \over64\pi^2} + {\l^2 T^5\over1024\pi m^3} - {9\l^2 T^3\over
    1024\pi^3 m} - {\l^3 T^7\over8192\pi m^5} + {9\l^3 T^5\over8192\pi^3
    m^3} + \ord{\l^4,\g^1} \right]  \nonumber \Enda
where the leading piece was given in \puteq{lead}.  The fully-dressed
1-loop tadpole is
  \Eqa \lefteqn{ \tadlead +  \gtad{20}{30}{10}{\Czi \put(8,9){\blot}} =
    \tadlead + {\d M^2\over\d\p}\, {-TM\over8\pi} = \tadlead + \slp\times}
    \label{tmv} \\
  && \left[ {-Tm\over8\pi} - {\l T^3\over64\pi m} + {3\l T^2
    \over16\pi^2} + {\l^2 T^5\over1024\pi m^3} - {63\l^2 T^3\over
    1024\pi^3 m} - {\l^3 T^7\over8192\pi m^5} + {63\l^3 T^5\over8192\pi^3
    m^3} + \ord{\l^4,\g^1} \right]  \nonumber \Enda
The difference is
  \Eq \Delta\equiv \gtad{20}{30}{10}{\Czi \put(8,9){\blot}} -
    \gtad{20}{30}{10}{\Czi} = \slp\left[ {9\l T^2\over
    64\pi^2} - {27\l^2 T^3\over 512\pi^3 m} + {27\l^3 T^5\over4096\pi^3
    m^3} + \ord{\l^4,\g^1} \right] \Endl{delta}
The interpretation of $\Delta$ as a miscounting of graphs
will become apparent after we compute the relevant Feynman diagrams.

\subsection{The Diagrams}

Besides the leading result \puteq{lead}, the diagrams explicitly give
  \Eq \tad{2}{20}{30}{10}{\Cz} = \slp {-Tm\over8\pi} \ab{-1}1 \Endl{dga}
  \Eq \tad{4}{20}{40}{10}{\Cz\MU\Cn} = \slp {-\l T^3\over64\pi m} \ab01
    \Endl{dgb}
  \Eq \tad{8}{40}{30}{20}{\Cz\MR\Cn\ML\ML\Cn} = \slp {\l^2 T^5\over1024\pi
    m^3} \ab11 \Endl{dgc}
  \Eq \tad{16}{40}{40}{20}{\Cz\MR\Cn\ML\ML\Cn\MR\MU\Cn} = \slp {-\l^3 T^7
    \over8192\pi m^5} \ab21 \Endl{dgd}
  \Eq \tad{4}{20}{50}{10}{\Cz\Mu\Cz} = \slp {3\l T^2\over64\pi^2} \ab{-1}2
    \Endl{dge}
  \Eq \tad{8}{20}{60}{10}{\Cz\Mu\Cz\MU\Cn} +
      \tad{4}{30}{50}{10}{\Cz\MR\Cn\ML\Mu\Cz} = 0 \ab02 \Endl{dgf}
  \Eq \tad{16}{35}{50}{15}{\Cz\Mu\Cz\MR\Cn\ML\ML\Cn} +
      \tad{16}{40}{50}{20}{\Cz\MR\Cn\ML\ML\Cn\MR\Mu\Cz} +
      \tad{ 8}{50}{40}{30}{\Cz\Ml\Cz\Mr\MU\Cn\MD\MR\Cn} +
      \tad{ 8}{30}{60}{10}{\Cz\MR\Cn\ML\Mu\Cz\MU\Cn} = 0 \ab12 \Endl{dgg}
  \Eq \tad{8}{20}{70}{10}{\Cz\Mu\Cz\Mu\Cz} +
      \tad{8}{60}{30}{30}{\Cz\Mr\Cz\Ml\Ml\Cz} = \slp {-9\l^2
    T^3\over1024 \pi^3 m} \ab{-1}3 \Endl{dgh}
  \Eqa
    \tad{16}{20}{80}{10}{\Cz\Mu\Cz\Mu\Cz\MU\Cn} \!&\!+\!&\!
    \tad{ 8}{30}{70}{10}{\Cz\Mu\Cz\MR\Cn\ML\Mu\Cz} \!\!+
    \tad{ 8}{30}{70}{10}{\Cz\MR\Cn\ML\Mu\Cz\Mu\Cz} +
    \tad{ 8}{40}{60}{30}{\Cz\Ml\Cz\Mr\Mu\Cz\MU\Cn} +
    \tad{ 8}{50}{50}{30}{\Cz\Ml\Cz\Mr\MR\Cn\ML\Mu\Cz} +
    \tad{16}{60}{40}{30}{\Cz\Ml\Cz\Mr\Mr\Cz\Ml\MU\Cn} \nonumber\\
    &=& \slp {9\l^3 T^5\over 8192\pi^3 m^3} \ab03 \Endla{dgi}
  \Eq \tad{16}{20}{90}{10}{\Cz\Mu\Cz\Mu\Cz\Mu\Cz} +
      \tad{16}{60}{50}{30}{\Cz\Mu\Cz\Mr\Cz\Ml\Ml\Cz} +
      \tad{ 8}{40}{70}{10}{\Cz\Mr\Cz\Ml\Mu\Cz\Mu\Cz} +
      \tad{16}{60}{50}{30}{\Cz\Ml\Cz\Mr\Mr\Cz\Ml\Mu\Cz} = 0 \ab{-1}4
    \Endl{dgj}
Note that the order in $\b$ is the number
of zero-mode loops, and the order in $\a$ is the number of non-zero-mode
loops minus one.

A somewhat surprising result is that, except for the figure eight daisy graph
in \puteq{dge}, all $\ord{\b^2}$ contributions sum to zero.  That all
$\ord{\b^2}$ contributions from daisy and superdaisy diagrams cancel for
any number of loops can be seen by schematically writing the solution to
the quadratic gap equation [\ie\ ignoring the last term in \puteq{gap}] as
  \Eq M \sim \b^2 + \b \sqrt{1 + \b^2} \Endl{gapquad}
The first term of $\ord{\b^2}$ corresponds to the subleading daisy graph of
\puteq{dge}, which cannot cancel with a superdaisy because superdaisies
first occur at three loops.  Since all other terms in $M$ are odd in $\b$,
the $\ord{\b^2}$ contributions to the effective potential arising from zero
modes (which at the tadpole level are proportional to $M$) must cancel.
This remains true only while the trilinear coupling can be ignored.  When
they are included, $M$ is the solution of a cubic equation containing both
even and odd powers of $\b$.  This result has implications for the
electroweak theory, where gauge boson gap equations can be approximated by
a quadratic \cite{us2}.\footnote{Subsequent works have explored the
electroweak gap equations in more detail \cite{eqz2,buch}, and are in
agreement with \cite{us2} up to, but not including, the $\phi$-independent
magnetic mass which is $\sim g^2 T$. The effect of the magnetic mass term
on the potential is $\ord{g^5}$ for $\phi \sim T$, and hence subleading in
a consistent $\ord{\b^2}$ calculation. However, at smaller values of $\phi$
the magnetic mass becomes increasingly important.}

Comparing the individual diagrams with the expansions \puteq{tm} and
\puteq{tmv}, we see that the partially-dressed tadpole gives
precisely the correct results.  Full dressing, \puteq{tmv}, leads to 3
erroneous terms, starting at $\ord{\b^2}$.  Graphical and algebraic
iteration of the gap equations show that one
third of $\Delta$ in \puteq{delta} arises from an overcounting of diagrams
in eqs.~(\ref{dge}, \ref{dgh}, and \ref{dgi}) by two, three, and three,
respectively.  Full dressing overcounts the individual superdaisies and
subleading daisies by a common factor, so that the sum still vanishes.
However, this cancellation is no longer possible when trilinear couplings
are reintroduced.

\subsection{Lollipops} \label{lolsec}

The other two thirds of $\Delta$ arise from an attempt to include the
lollipop (and its dressed cousins), which is also super-daisy order
\cite{us2,deriv,parw}:
  \Eq \tad{6}{20}{30}{10}{\Cai \put(9,10){\line(0,1){20}}
    \put(11,10){\line(0,1){20}}} = \slp {\l T^2\over32 \pi^2} \left[
    \ln\left( M^2\over T\mub \right) + 1.65 \right] \ab{-1}{2} \Endl{pop}
where $\mub$ is a renormalization scale, often taken to be $T$; and $M$
is the improved mass of \puteq{Mimp}.  The infrared behavior is calculable
just from the zero modes.  Recall that $m/T \sim \sqrt\l$, so for any
reasonable Higgs mass the log term is near unity.

In the partial dressing method the lollipop is not considered a
``daisy-type'' diagram; the result \puteq{pop} is just added to $V'_{pd}$
to give $V'_{pd+l}$.

The full dressing method sees this diagram as an improved-propagator main
loop attached to a vertex dressed with an improved-propagator bubble.
Algebraically, it arises from
  \Eq \frac12 {\d M^2\over\d\p} I_1(M^2) \ni
     \frac12 \left[c_3 c_4 I_2(M^2) \right] I_1(M^2) \Endl{plol}
Symmetry factors of $\frac12$ from the main loop and $\frac12$ from the
vertex loop combine with a factor of 2 ways to attach the external line to
the 4-point vertex, giving an overall factor of $\frac12$ instead of the
correct $\frac16$.  Overlapping momenta and non-zero modes are ignored.
Then
  \Eq \mfrac12 (-6\l\p)(-6\l) \left( -Tm\over4\pi \right) \left( T\over
    8\pi m \right) = \slp {3\l T^2\over 32\pi^2} \Endl{eqzlol}
which is the leading term of $\frac23\Delta$.  Because overlapping momenta
are ignored, the logs of the true calculation are not reproduced.  In
principle, using the momentum-dependent self energy in the gap equations
would result in inclusion of the logs, but the combinatoric miscounting
would remain.

The leading term of $V'_{fd} - V'_{pd+l}$ (\ie\ the error in the
full-dressing calculation) arises from one extra figure-eight tadpole
[\puteq{dge}] and two extra lollipops [$2/3$ of \puteq{eqzlol} in the
approximation that overlapping momenta are ignored]; these are then
subtracted off with
  \Eq V'_{comb} = \slp {-7\l T^2\over 64\pi^2} \Endl{Vcomb}
as given in \cite{eqz2} (but apparently neglected in \cite{eqz1}).  At each
order in $\b$ more terms of $V'_{comb}$ would need to be calculated to
correct the full-dressing method, since [as we saw in \puteq{fdress} and
\puteq{delta}] full dressing overcounts an infinite class of diagrams.  The
CJT technique we will discuss in Sec.~4 provides a systematic, if
cumbersome, way to calculate $V'_{comb}$.

\section{Tying Up Loose Ends}

Many approximations were made in the previous section in order to
facilitate an explicit counting of diagrams.  Here we will re-examine them
and develop a general $\ord{\b^2}$ procedure for calculating $V'_{pd+l}$.

\subsection{Non-Daisies}

Diagrams besides the ``daisy-type'' ones and the lollipops are all either
higher-order in $\g$ or at least $\ord{\b^3}$, \eg
  \Eq \tad{12}{20}{40}{10}{\Ca\MU\MU\Md\Ca} \sim \slp {3\l^2 T^3\over128
    \pi^3 m} \ln(T/m) \ab{-1}3 \End	
(The infrared behavior was calculated from zero modes as for the lollipop.)
There would be little point deriving a potential accurate to $\ord{\b^3}$
unless these diagrams were also included.

\subsection{Log Terms and Dressed Non-Zero-Mode Loops}

By ignoring the ellipsis in \puteq{vac}, we not only reduced the number of
diagrams to calculate, but also evaded the question of whether to dress
non-zero-mode loops.  In Table~\ref{tab2} we now restore terms proportional
to $L\equiv \ln(\mub^2/T^2)-2c_B$, where $2c_B = 2\ln(4\pi) - 2\gamma_E
\approx 3.9076$ and $\mub$ is a renormalization scale.  When determining the
order of a diagram we will treat $L$ as $\ord{1}$.

  \begin{table} \centering \begin{tabular}{|r|cccc|} \hline
  integral: & $I_0$ & $I_1$ & $I_2$ & $I_3$ \\ \hline
  non-zero modes: &
    $\left( \frac{\pi^2 T^4}{45} \!-\! \frac{T^2 m^2}{12} \!+\! \frac{m^4
      L}{32\pi^2} \right)$ &
    $\left( \frac{T^2}{12} \!-\! \frac{m^2 L}{16\pi^2} \right)$ &
    $\frac{L}{16\pi^2}$ & $0$ \rule[-8pt]{0pt}{23pt}\\
  zero modes: & $\frac{T m^3}{6\pi}$ & $\frac{-Tm}{4\pi}$ & $\frac{T}{8\pi
    m}$ & $\frac{T}{32\pi m^3}$ \rule[-8pt]{0pt}{23pt}\\ \hline
  \end{tabular} \caption{Rules including log term $L\equiv
    \ln(\mub^2/T^2) - 3.9076$} \label{tab2} \end{table}

The diagrams of eqs.~(\ref{dga}-\ref{dgj}) now have subleading pieces, and
new diagrams (with non-zero-mode loops of several propagators) appear.  We
will spare the reader by mentioning just the two new $\ord{\b^2}$
contributions:
  \Eq \mfrac12 \hspace{2pt} \core{-5}{10}{20}{5}{15}{
    \put(5,0){\line(0,1){10}}\Cn} \,\hbox{[subleading]} = \slp {-m^2 L\over
    32\pi^2} \ab{-2}{2} \Endl{sublead}
  \Eq \tad{4}{10}{30}{5}{\Md\MU\Cn\MU\MU\Md\Cn} = \slp {-\l T^2 L\over128
    \pi^2} \ab{-1}{2} \Endl{CnCn}
These (and higher order generalizations) can be accounted for by keeping
the $L$-term in the 1-loop tadpole, \puteq{lead}, and using the improved
mass $M$ in the $\frac{m}T$ expansion.  This corresponds to improving both
zero- and non-zero-mode propagators as done in \cite{us2}.  The full
dressing method of references \cite{eqz1,eqz2} improves only zero modes, and
therefore omits the $\ord{\b^2}$ graph in \puteq{CnCn}.

\subsection{3-Point Vertices}

Let us now examine $\ord{\g^1}$ diagrams containing three 3-point vertices
and up to 3 loops.  When $\lambda \p^2$ terms are retained in the gap
equation \puteq{gap}, results \puteq{tm} and \puteq{tmv} are modified to
  \Eq \tad{2}{20}{30}{10}{\Cai} = \slp \left[ {\rm(old)} + {9 \l^2 \p^2
    T^2\over64\pi^2 m^2} - {9\l^3 \p^2 T^4\over 256\pi^2 m^4} + {27\l^3
    \p^2 T^3\over 512\pi^3 m^3} + \ord{\l^4,\g^2} \right] \Endl{t3m}
  \Eq \tad{2}{20}{30}{10}{\Cai \put(8,9){\blot}} = \slp \left[
    {\rm(old)} + {81\l^3 \p^2 T^3\over 512\pi^3 m^3} + \ord{\l^4,\g^2}
    \right] \Endl{t3mv}

The new 2-loop diagram is the ``setting sun'' tadpole
  \Eq \tad{4}{20}{30}{10}{\Ca \put(0,20){\line(1,0){20}}} = \slp
    {6\l^2 \p^2 T^2\over64\pi^2 m^2} \;{\rm (true)}, \quad \approx \slp
    {9\l^2 \p^2 T^2\over64\pi^2 m^2} \;{\rm (naive)} \abg021 \Endl{dercor}
The ``true'' result is from the double integral done properly; the
``naive'' result comes from ignoring the overlapping momenta and using
Table~\ref{tab1}, assigning two propagators to each of the integrals.
This is the approximation that has been criticized in \cite{deriv},
and amounts to approximating a momentum-dependent self energy $\Pi(Q^2)$
by its zero momentum value $\Pi(0)$.  The naive result is $3/2$ times the
true result, which as suggested in \cite{deriv} is a significant error.
This error is exacerbated in the electroweak theory, where logarithms
from analogous diagrams are lost if $\Pi(0)$ is used.
Here, we are interested in counting arguments which are independent of
whether one uses $\Pi(0)$ or $\Pi(Q^2)$.

Note that partial dressing of the 1-loop tadpole [\puteq{t3m}] correctly
reproduces the ``naive'' result, while full dressing [\puteq{t3mv}] does
not.  More subtly, full dressing counts the setting sun tadpole once as a
dressed propagator and twice as a dressed vertex, which happen to cancel
(because overlapping momenta are treated differently) and give zero.  If
$\Pi(Q^2)$ were used, the diagrams would instead add, leading to a miscount
of three.\footnote{In the electroweak calculation of \cite{eqz2}, the gauge
boson analogue of the figure eight graph \puteq{dge} was subtracted once,
and the gauge boson analogue of the setting sun graph \puteq{dercor} was
subtracted twice by a term $V_{comb}$, correctly compensating for the
miscounting of these diagrams. However, topologically equivalent diagrams
outside the pure gauge sector remain overcounted (for example, the setting
sun vacuum graph consisting of one Higgs and two gauge boson propagators).}

At 3 loops we have
  \Eq \tad{4}{20}{40}{10}{\Ca\MU\Cn \put(0,20){\line(1,0){20}}} +
    \tad{4}{30}{30}{10}{\Ca\MR\Cn \put(2.93,12.93){\line(1,1){14.14}}}
    = \slp {-6 \l^3 \p^2 T^4\over256\pi^2 m^4} \;{\rm (true)}, \quad
    \approx \slp {-9 \l^3 \p^2 T^4\over256\pi^2 m^4} \;{\rm (naive)}
    \abg121 \End
  \Eqa \lefteqn{\tad{4}{20}{50}{10}{\Ca\Mu\Cz \put(0,20){\line(1,0){20}}}
    + \tad{4}{20}{50}{10}{\Cz\Mu\Ca \put(0,40){\line(1,0){20}}} +
    \tad{4}{40}{30}{10}{\Ca\Mr\Cz \put(2.93,12.93){\line(1,1){14.14}}}} \\
    && = \slp {18 \l^3 \p^2 T^3\over512\pi^3 m^3} \;{\rm (true)}, \quad
    \approx \slp {27 \l^3 \p^2 T^3\over512\pi^3 m^3} \;{\rm (naive)}
    \abg031 \nonumber \Enda
We see again that partial dressing correctly counts the naive calculations
of these graphs.  Since the naive results are again $3/2$ times the true
results (hard loop dressings do not affect momentum flow), we can
multiply the last term in the gap equation (\ref{gap}) by $2/3$ to correct
for using $\Pi(0)$ instead of $\Pi(Q^2)$.  This is just a simple way to
implement our explicit calculations, and does not represent a systematic
improvement of the partial dressing method.

We can show that no tadpole graphs with $j_\g > 1$ contribute at
$\ord{\b^2}$.  Roughly, every additional factor of $\g$ means two more
3-point vertices, which form either a zero-mode loop (contributing $\b$) or
a 2-propagator non-zero-mode loop (contributing $\b^2$).  More precisely, a
graph with $Z$ zero-mode loops, $N_1$ 1-propagator non-zero-mode loops,
$N_2$ 2-propagator non-zero-mode loops, $f$ 4-point vertices, and $t$
3-point vertices, obeys
  \Eq f + \frac12 (t+1) = Z+N_1+N_2, \qquad j_\a-j_\g+1 = N_1-N_2, \qquad
    j_\b = Z+2N_2, \qquad j_\g = \frac12 (t-1) \End
Except for the leading diagram \puteq{lead}, $f\ge N_1$, from which it
follows that
  \Eq j_\b \ge (Z+N_2) \ge \frac12 (t+1) = j_\g+1 \End

\subsection{The Full Result of Partial Dressing}

To summarize, our results, good to $\ord{\b^2}$ (and all orders in $\a$ and
$\g$), are:
  \Eq V'_{pd+l} = \tad{2}{20}{30}{10}{\Cai} +
    \tad{6}{20}{30}{10}{\Cai \put(9,10){\line(0,1){20}}
      \put(11,10){\line(0,1){20}}}
    = \slp \left[ {T^2\over24} - {TM\over8\pi} - {M^2 L\over32\pi^2} +
      {\l T^2\over32\pi^2} \left\{ \ln\left( M^2\over T\mub \right) + 1.65
      \right\} \right],  \Endl{finale}
  $$ M^2 = m^2 + {\l T^2\over4} - {3\l TM\over4\pi} - {2\over3}\cdot
    {9\l^2\p^2 T\over 4\pi M}, \qquad L\equiv \ln(\mub^2/T^2) - 3.9076 $$
Here the ``true'' results of \puteq{dercor} {\it etc.} have been
incorporated by the new factor of $2/3$ in the gap equation.  Note that
despite the resummation of non-zero modes, no linear term has been
generated, since $V'_{pd+l}(\p=0) = 0$.

In contrast, even with the counting correction \puteq{Vcomb},
$V'_{fd+comb}$ is still incorrect at $\ord{\b^2}$ (even for $\g\ll 1$), due
to overlapping momenta in the lollipop and the omission of \puteq{CnCn}.

\section{Other Methods}

\subsection{The CJT Technique}

Amelino-Camelia and Pi \cite{acp} employ the technique of Cornwall, Jackiw
and Tomboulis (CJT) \cite{cjt} to derive an effective action
  \Eq \Gamma[\p] = I_{\rm cl}[\p] + \Gamma^{(1)}[\p,G] + \Gamma^{(2)}[\p,G]
    - {\rm Tr}\left\{ {\delta\Gamma^{(2)}[\p,G] \over \delta G} G \right\}
    \Endl{cjt1}
where $\Gamma^{(1)}$ is the improved 1-loop vacuum graph, $\Gamma^{(2)}$
consists of 2PI graphs with improved propagators and unimproved
vertices, and the improved propagator $G$ comes from a gap equation
  \Eq G^{-1} = D^{-1} + 2 {\delta\Gamma^{(2)}[\p,G] \over \delta G} \End
Roughly translated, \puteq{cjt1} says that any $n$-propagator diagram arising
from the fully dressed 1-loop vacuum graph must be subtracted off $(n\!-\!1)$
times.

Suppose we put only the (2-propagator) figure-eight vacuum graph of
\puteq{tad} into $\Gamma^{(2)}$ (as done in \cite{acp}). The gap equation
becomes
  \Eq \Bp(32,3) \put(0,0){\line(1,0){32}} \put(0,3){\line(1,0){32}} \Ep
    \;=\; \Bp(32,1) \put(0,1){\line(1,0){32}} \Ep \;+\;
    \core{0}{32}{20}{16}{10}{\Cai \put(0,0){\line(1,0){32}}} \qquad
  \bM^2 = m^2 - \frac12 c_4 I_1(\bM^2) \Endl{bMdef}
and the tadpole equivalent of \puteq{cjt1} is
  \Eq V'_{cjt1} = \tad{2}{20}{30}{10}{\Cai \put(8,9){\blot}} -
      \tad{4}{20}{50}{10}{\Cai\Mu\Cai \put(8,9){\blot}} =
      \mfrac12 C_3\, I_1(\bM^2) + \frac14 C_3 c_4\, I_1(\bM^2) I_2(\bM^2)
      \End
But the gap equation implies the identity
   \Eq  C_3 [ 1 - \frac12 c_4 I_2(\bM^2) ] = c_3 \End
so
  \Eq V'_{cjt1} = \mfrac12 c_3\, I_1(\bM^2) = \tad{2}{20}{30}{10}{\Cai} =
    V'_{pd} \End
The CJT procedure with this $\Gamma^{(2)}$ gives precisely the same result
as the partially-dressed tadpole of \puteq{tm}, in the limit $\g\ll 1$.
We have already noted that the lollipop is leading order in $\g$ and must
be added by hand to $V'_{pd}$, so the same applies to this version of CJT.

Now let us also include the (3-propagator) 2-loop ``setting sun'' diagram
in $\Gamma^{(2)}$.  The gap equation is \puteq{mgap}, and
  \Eq V'_{cjt2} = \tad{2}{20}{30}{10}{\Cai \put(8,9){\blot}}
    -  \tad{4}{20}{50}{10}{\Cai\Mu\Cai \put(8,9){\blot}}
    -2\, \tad{6}{20}{30}{10}{\Cai \put(9,10){\line(0,1){20}}
       \put(11,10){\line(0,1){20}}}
    -2\, \tad{4}{20}{30}{10}{\Cai \put(0,19){\line(1,0){20}}
       \put(0,21){\line(1,0){20}} \put(8,9){\blot}} \Endl{cjt2}
The gap equation now implies the identity
   \Eq  C_3 [ 1 - \frac12 c_4 I_2(M^2) - c_3^2 I_3(M^2)] =
          c_3  + c_3 c_4 I_2(M^2) \End
so that (ignoring overlapping momentum)
  \Eq V'_{cjt2} = \tad{2}{20}{30}{10}{\Cai} +
     \tad{6}{20}{30}{10}{\Cai \put(9,10){\line(0,1){20}}
       \put(11,10){\line(0,1){20}}} = V'_{pd+l} \Endl{cjt3}
The partially-dressed tadpole (for general $\g$) plus lollipop,
\puteq{finale}, has now been recovered.  We again see in \puteq{cjt2} that
full dressing overcounts 1 extra figure-eight, 2 extra lollipops, and 2
extra setting-suns; the CJT technique provides a systematic way of
calculating $V'_{comb}$.

When done more carefully, the CJT technique may be capable of handling
overlapping momenta, but we are unaware of any such analysis.

\subsection{Restoring The Loop Expansion}

Arnold and Espinosa \cite{deriv} suggest another method of resumming
daisies which restores the loop expansion.  They note that each zero-mode
loop costs at least a factor of $\b$, so to compute to $\ord{\b^2}$, one
need evaluate only graphs with two or fewer zero-mode loops.  This
avoids any combinatoric complications due to field-dependent mass shifts.
Hard thermal loops on zero-mode propagators are resummed by shifting the
mass with a temperature-dependent but field-independent quantity,
\Eq \Bp(32,3) \thk{\put(0,0){\line(1,0){32}} \put(0,3){\line(1,0){32}}} \Ep
    \;=\; \Bp(32,1) \put(0,1){\line(1,0){32}} \Ep \;+\;
    \Bp(32,10) \put(0,0){\line(1,0){32}} \put(16,6){\circle{10}} \Ep \qquad
  \bm^2 = m^2 + {\l T^2\over4} \End
so $m\to\bm$ only in the bottom row of Table \ref{tab2}.  A ``thermal
counterterm'' is introduced to cancel the overcounting of graphs which
occurs when improved propagators are used in a loop expansion \cite{parw}:
  \Eq \Bp(32,6) \thk{\put(0,1){\line(1,0){32}} \put(0,3){\line(1,0){32}}}
     \put(13,0){\rm x} \Ep \;=\; {\l T^2\over 4} \End
The counterterm ensures that the one-point function result remains
unchanged even if non-zero modes are also resummed.  Then
\Eqa V'_{ae} \!&\!=\!&\! \tad{2}{20}{30}{10}{\thk{\Cai}} +
  \tad{4}{20}{50}{10}{\thk{\Cai\Mu\Cai}}
   +   \tad{2}{20}{30}{10}{\thk{\Cai} \put(7,26){\rm x}}
  + \tad{6}{20}{30}{10}{\thk{\Cai \put(9,10){\line(0,1){20}}
    \put(11,10){\line(0,1){20}}}} +
  \tad{4}{20}{30}{10}{\thk{\Cai \put(0,19){\line(1,0){20}}
    \put(0,21){\line(1,0){20}}}} \label{aeres} \\
  \!&\!=\!&\! \slp \left[ {T^2\over24} - {T\bm\over8\pi} - {m^2
    L\over32\pi^2} - {\l T^2( L-6)\over 128\pi^2} +
    {\l T^2\over32\pi^2} \left\{ \ln\left( \bm^2\over T\mub \right) + 1.65
    \right\} + {6\l^2\p^2 T^2\over 64\pi^2 \bm^2} \right] \nonumber \Enda
which agrees with \puteq{finale} to $\ord{\b^2}$.

The result for diagram counting is identical to partial dressing.
However, because the two loop graphs are being explicitly evaluated,
overlapping momentum are always handled correctly.  This is a significant
improvement over the partial dressing method.

Another advantage to this zero-mode loop expansion is that it easily
generalizes to higher order in $\b$.  One must be careful, however, if it
becomes necessary to shift the mass in a field dependent way.  In the
Abelian Higgs model, a cancellation \cite{deriv} eliminates the need to do
this at $\ord{\b^2}$.  It is not clear to us if this will be true at
$\ord{\b^3}$.  If not, it is important to partially dress rather than to
simply insert the improved mass into the one loop vacuum graph.

\subsection{The Two-Point Way}

Near the completion of our work, we became aware of another treatment of
the electroweak phase transition by Buchmuller, Fodor, Helbig, and Walliser
\cite{buch}. These authors solve gap equations for scalar and vector boson
propagators (2-point functions), and integrate (effectively, twice) to get
the effective potential.  As they point out, the result contains all of the
$\ord{\b}$ corrections, but only some of the $\ord{\b^2}$ corrections.
Applied to scalar $\p^4$ theory, we believe their procedure is equivalent
to integrating our gap equation (\ref{mgap}) twice. This differs from
the partial dressing method, which inserts the solution of the gap equation
into a one loop tadpole, going one iteration further in the improvement
of the effective potential.

By differentiating \puteq{finale} and using \puteq{vga}, we see that our
$\ord{\b^2}$, partial-dressing improved mass-squared is
  \Eqa \left( V_0'' + V''_{pd+l} \right)^{-1} \!&\!=\!&\!
    \Bp(32,1) \put(0,1){\line(1,0){32}} \Ep \;+\;
    \core{0}{32}{20}{16}{10}{\Cai \put(0,0){\line(1,0){32}}} \;+\;
    \core{-10}{32}{20}{16}{10}{\put(0,10){\line(1,0){6}} \Cai
      \put(26,10){\line(1,0){6}}} \nonumber\\
    &+& \core{-10}{52}{20}{16}{10}{\put(0,10){\line(1,0){6}} \Cai\Mr\Cai
      \put(43,8){\blot} \put(46,10){\line(1,0){6}}} \;+\;
    2\, \core{-10}{32}{20}{16}{10}{\put(0,20){\line(1,0){16}} \Cai
      \put(6,9){\line(1,0){20}} \put(6,11){\line(1,0){20}}
      \put(26,10){\line(1,0){6}}} \;+\;
    \core{-10}{32}{20}{16}{10}{\put(0,20){\line(1,0){16}} \Cai
      \put(6,9){\line(1,0){20}} \put(6,11){\line(1,0){20}}
      \put(26,10){\line(1,0){6}} \put(14,17){\blot}} \;+\;
    \core{-10}{32}{20}{16}{10}{\put(0,10){\line(1,0){6}} \Cai
      \put(26,10){\line(1,0){6}} \put(15,0){\line(0,1){20}}
      \put(17,0){\line(0,1){20}} \put(23,8){\blot}} \;+\;
    \core{-10}{32}{20}{16}{10}{\put(0,10){\line(1,0){6}} \Cai
      \put(26,10){\line(1,0){6}} \put(6,9){\line(1,0){20}}
      \put(6,11){\line(1,0){20}}} \Endla{Vpp}
which contrasts sharply with the $M^2$ of \puteq{mgap}; to be precise, the
2-point method of \cite{buch} misses all the 2-loop 1PI diagrams of
\puteq{Vpp}.  In addition, it suffers the usual problems with overlapping
momenta.

\subsection{The Hybrid Way}

We have seen that partial dressing makes correct counting easy, but
overlapping momenta [in the last term of the gap equation \puteq{gap}] are
problematic.  We now propose using the gap equation of \puteq{bMdef}, which
dresses the propagator with only momentum-independent loops, and calculating
the setting sun vacuum graph (which gives both the setting sun tadpole and
the lollipop) separately, with only hard thermal loop dressings, as done by
Arnold and Espinosa.  Then we get a potential correct to $\ord{\b^2}$ from
only two graphs (and one quadratic gap equation):
  \Eqa V'_{hyb} &=& \tad{2}{20}{30}{10}{\Cai} + {\d\over\d\p} \left[
    \vac{12}{20}{20}{10}{\thk{\Cai \put(0,9){\line(1,0){20}}
      \put(0,11){\line(1,0){20}}}} \right] \label{hybres} \\
  \!&\!=\!&\! \slp \left[ {T^2\over24} - {T\bM\over8\pi} - {\bM^2 L\over
    32\pi^2} \right] + {\d\over\d\p} \left[ {3\l^2\p^2 T^2\over32\pi^2}
    \left\{ \ln\left( \bm^2\over T\mub \right) + 1.65 \right\} \right]
    \nonumber \Enda
  \Eqa \Bp(32,3) \put(0,0){\line(1,0){32}} \put(0,3){\line(1,0){32}} \Ep
    \;=\; \Bp(32,1) \put(0,1){\line(1,0){32}} \Ep \;+\;
    \core{0}{32}{20}{16}{10}{\Cai \put(0,0){\line(1,0){32}}} &\quad&
    \bM^2 = m^2 + {\l T^2\over4} - {3\l T \bM\over4\pi}, \nonumber\\
  \Bp(32,3) \thk{\put(0,0){\line(1,0){32}} \put(0,3){\line(1,0){32}}} \Ep
    \;=\; \Bp(32,1) \put(0,1){\line(1,0){32}} \Ep \;+\;
    \Bp(32,10) \put(0,0){\line(1,0){32}} \put(16,6){\circle{10}} \Ep
    &\quad& \bm^2 = m^2 + {\l T^2\over4}  \nonumber\Enda
and again $L\equiv \ln(\mub^2/T^2) - 3.9076$.
This hybrid method generalizes easily to $\ord{\b^3}$ just by adding all
3-loop vacuum graphs with overlapping momenta, and should be just as
applicable to more complicated theories such as the Standard Model.
We expect the computational utility of the hybrid method to be more
apparent in such generalizations.

\section{Conclusion: What's Hot and What's Not}

\subsection{Summary of Results}

We have examined various prescriptions for calculating $\ord{\b^2}$
contributions to the effective potential in a scalar $\p^4$ theory, by
comparing the first few terms in a loop expansion to explicit Feynman graph
computations.

We showed that fully-dressed tadpoles (or equivalently, dressed vacuum
diagrams) overcount an infinite class of diagrams, overcount and
incorrectly calculate lollipop-type diagrams, miss significant
contributions arising from non-zero modes, and suffer corrections due to
overlapping momenta [approximating $\Pi(Q^2) \approx \Pi(0)$].  In order to
calculate $V'$ to $\ord{\b^2}$ correctly, one needs to subtract the
overcounted figure-eight tadpole and lollipop (\ie\ include $V'_{comb}$),
compensate for overlapping momentum corrections (in the lollipop), restore
the hard-loop dressed setting sun tadpole, and include the hard
figure-eight of \puteq{CnCn}.

Partially-dressed tadpoles completely miss lollipop-type diagrams, and
suffer overlapping momentum corrections.  In order to calculate $V'$ to
$\ord{\b^2}$ correctly, one needs to add the lollipop by hand (as done in
\cite{us2}), and compensate for momentum corrections in the hard-loop
dressed setting sun tadpole [as seen in \puteq{dercor}].  The prescription
for scalar $\p^4$ theory is given in \puteq{finale}.

The CJT method, \puteq{cjt1}, provides a systematic way of removing the
overcounted diagrams of the full dressing method, but we do not know how to
extend it to correctly calculate overlapping momenta.  As it stands now, it
is equivalent to partial dressing.

The hard-loop dressing of Arnold and Espinosa, \puteq{aeres}, counts
diagrams correctly (through the use of ``thermal counterterms''), and no
overlapping momentum errors are incurred because all such diagrams are
calculated explicitly.  This task is somewhat easier if one sticks to
vacuum graphs.

The 2-point method of Buchmuller \etal\ \cite{buch} does not seem to be
an attempt at a complete $\ord{\b^2}$ calculation.

Finally, we suggested in \puteq{hybres} a simple synthesis of the above
procedures.  A tadpole is partially dressed with only momentum-independent
loops (both zero and non-zero modes), and all other diagrams are calculated
by hand at the vacuum level, using hard-loop dressing.  At $\ord{\b^2}$ there
is only one such diagram, the setting sun.

\subsection{Outlook for the EWPT}

Although the analysis presented here is in the context of scalar $\p^4$
theory, the conclusions are equally valid for the electroweak phase
transition (the main difference being an exacerbation of errors due to
new graphs involving gauge bosons).  This allows us to examine recent
conflicting claims about the nature of the EWPT.

In a previous paper \cite{us2}, the authors, using partial dressing,
found $\ord{\b^2}$ contributions to the effective potential which weakened
the phase transition.  The transition remained first order over the range of
validity of our calculation.  We estimated the effects of ignoring
overlapping momentum, suggesting it would be numerically small.  It has
since been shown that this $\ord{\b^2}$ contribution is logarithmically
enhanced \cite{deriv}, so that the partial dressing method in \cite{us2} is
incomplete.  In particular, setting sun type diagrams need to be handled
more carefully to produce an effective potential reliable to $\ord{\b^2}$.

Espinosa, Quiros, and Zwirner \cite{eqz2}, using full dressing with a
$V_{comb}$ correction, find a weakened EWPT which becomes second order near
the limits of their range of validity.  They also ignore overlapping
momenta.  In addition, their $V_{comb}$ neglects some overcounted graphs in
the Higgs-gauge sector, and they ignore $\ord{\b^2}$ contributions
arising from non-zero-mode figure-eight graphs. For these reasons, their
results are suspect.

A more strongly first order EWPT has been reported by Arnold and Espinosa
\cite{deriv}.  We have seen that for $\p^4$ theory, their counting agrees
with partial dressing, and their method handles overlapping momenta
correctly, so we believe this result is reliable.  We have verified their
explicit computations only for the scalar $\p^4$ theory.  This method
seems easily generalizable to higher order in $\b$, though any such
generalization must take care to count graphs correctly if diagrams are
resummed in a field-dependent manner.  We expect the hybrid method of
\puteq{hybres} applied to the EWPT would give results similar to those of
Arnold and Espinosa.

\vskip.2in {\centering\large\bf Acknowledgments}

We thank Peter Arnold, Stamatis Vokos, Jose Ramon Espinosa, Mariano Quiros,
and Fabio Zwirner for their patient and helpful discussions.  CGB
acknowledges support from the National Science Foundation under grant
NSF-PHY-91-23780 and the U.S. Department of Energy under grant
DEFG-02-90ER40560.  DEB acknowledges support from the U.S. Department of
Energy under grant DEAC-03-81ER40050.  SDH acknowledges support from the
National Science Foundation under grant NSF-PHY-92-18167, the state of
Texas under grant TNRLC-RGFY9106, the Milton Fund of the Harvard Medical
School and from the Harvard Society of Fellows.


\newpage
\def\np#1{{\it Nucl. Phys.\ }{\bf #1}}
\def\pl#1{{\it Phys. Lett.\ }{\bf #1}}
\def\pr#1{{\it Phys. Rev.\ }{\bf #1}}

\def\baselinestretch{.8}

\end{document}